# Study of atomic entanglement in multimode cavity optics

Papri Saha[1], N. Nayak[2], and A. S. Majumdar[2]
[1]Department of Computer Science, Derozio Memorial College, Rajarhat Road, Kolkata-700136, India
[2]S.N.Bose National Centre for Basic Sciences, Block-JD, Sector-3, Salt Lake City, Kolkata-700098, India

**Abstract.** The resonant interaction between two two-level atoms and m- electromagnetic modes in a cavity is considered. Entanglement dynamics between two atoms is examined. In particular, we compare dynamical variations for different cavity modes as well as for different cavity photon numbers. The collapse and revival of entanglement is exhibited by varying the atom-photon interaction times.

## I. Introduction

The ability to entangle large numbers of atoms that interact with the electromagnetic field for long periods of time has received a great deal of attention in recent years [1-6]. Quantum dynamics of two-level systems (qubits), such as spins in a magnetic field, Rydberg atoms or Cooper Pair Boxes, coupled to a single mode of an electromagnetic cavity are of considerable interest in connection with NMR studies of atomic nuclei [7], Cavity Quantum Electrodynamics [8] and Quantum Computing [9] respectively. The simplest model, which captures the salient features of the relevant physics in these fields, is the Jaynes-Cummings model (JCM) [10] for the one-qubit case and its generalization for multi qubit systems known as Tavis and Cummings model (TCM) [11].

Until recently, the quantum light-atom entanglement has been almost exclusively considered at the level of a single mode-single atom interaction which is a natural setting for qubit-type quantum information processing, but an efficient interface between propagating light beams and atoms is yet to be demonstrated. Theoretical efforts have been devoted to control the entanglement because it has been realized that the controlled manipulation of such resources has its practical importance in the actual quantum information processing. Bipartite entanglement in the context of JCM has been widely studied [13-15]. The entanglement sharing in the two-atom TCM coupled to a single mode of the electromagnetic field has been described by various authors [12].

All the above works deal with a single mode cavity. Although they are widely studied models, confining the photons to a single mode of the cavity can be very difficult experimentally. Hence, study of bipartite entanglement in TCM in a multimode cavity is not out of place. In a recent paper, one of us [16] has analyzed the so-called 'collapse and revival' in the atomic population of a single atom in a multimode cavity. It has been shown there some interesting deviation of the atomic population revival times occur compared to a single-mode dynamics. This indicates that the bipartite entanglement in TCM in a multimode cavity could show a different dynamics compared to the single-mode case [17]. In this paper we wish to focus on the dynamics of entanglement between the atoms (or qubits) in m-mode cavity field.

This paper is organized as follows: A general introduction to the Hamiltonian of our system along with the parameters used to study population inversion and atomic entanglement is provided in Section II. In Sec III, entanglement dynamics of a bipartite system in a single mode field will be studied. Here we also compare the atomic population dynamics between single atom and two-atom case. In Section IV we study the entanglement dynamics for the m-mode radiation field. Finally, we summarize our results in Section V.

## II. Atom-field interaction Hamiltonian

The Tavis-Cummings model (TCM) [11] describes the simplest fundamental interaction between a single mode of the quantized electro-magnetic field and a collection of $l$ atoms under the usual two-level and rotating wave approximations. The two- atom ($l$=2) TCM (lower state $|b\rangle$, upper state $|a\rangle$, transition frequency $\omega$) can be represented by the spin operators

$$S_z = \sum_{j=1}^{l} \frac{\left[\left(|a\rangle\langle a|\right)_j - \left(|b\rangle\langle b|\right)_j\right]}{2}, \; S_+ = \left(|a\rangle\langle b|\right)_1 + \left(|a\rangle\langle b|\right)_2, \; S_- = \left(|b\rangle\langle a|\right)_1 + \left(|b\rangle\langle a|\right)_2 \tag{1}$$

It follows that the resonant interaction of a $m-$mode quantized field with two two-level atoms is described by the Hamiltonian (in the interaction picture)

$$H = H_0 + H_{\text{int}}, \tag{2}$$

$$H_0 = \sum_{k=1}^{m} \hbar\omega a_k^+ a_k + \hbar\omega S_z, \tag{3}$$

$$H_{\text{int}} = \hbar \sum_{k=1}^{m} g_k \left(S_+ a_k + S_- a_k^+\right) \tag{4}$$

Let, $g_1 = g_2 = \ldots = g_m = g$ (for simplicity we have taken them to be equal).
Therefore,

$$H_{\text{int}} = \hbar g \left( S_+ \sum_{k=1}^{m} a_k + S_- \sum_{k=1}^{m} a_k^+ \right) \tag{5}$$

For the interaction of atoms with a quantized radiation field, the unitary time-evolution operator is given by

$$U(t) = \exp\left( \frac{-iH_{\text{int}} t}{\hbar} \right) \tag{6}$$

where the interaction picture Hamiltonian $H_{\text{int}}$ is given by Eq. (5). The atom-field state at time $t$ in terms of the state at time $t = 0$ is given by

$$|\psi(t)\rangle = U(t)|\psi(0)\rangle \tag{7}$$

We take atoms initially in the excited state $|a_1, a_2\rangle$ and the field as a superposition of number states, i.e.,

$$|\psi(0)\rangle = \sum_{n_1, n_2, \ldots, n_m} c_1(0) c_2(0) \ldots c_m(0) |a_1, a_2, n_1, n_2, \ldots, n_m\rangle, \tag{8}$$

where $n_m$ represents the number of photons in mode $m$.

The population dynamics of the atoms can be studied in terms of the function $W(t)$ defined as

$$W(t) = \sum_{n_1, n_2, \ldots, n_m} \left( \left|\langle a_1, a_2, n_1, n_2, \ldots, n_m |\psi(t)\rangle\right|^2 - \left|\langle b_1, b_2, n_1, n_2, \ldots, n_m |\psi(t)\rangle\right|^2 \right) \tag{9}$$

that gives the probability of finding both the atoms in the excited state minus the probability of atoms in the ground state.

We use the state $|\psi(t)\rangle$ to obtain the measure of entanglement for the two atoms. The *Concurrence* and *Entanglement of Formation* [18,19] for a bipartite density operator $\rho_{atom}$ is given by

$$E_F\left(\rho_{atom}\right)=h\left(\frac{1+\sqrt{1-C^2\left(\rho_{atom}\right)}}{2}\right) \tag{10}$$

For the wave function $\left|\psi(t)\right\rangle$ of atom-field system, the two-atom density operator $\rho_{atom}(t)$ is obtained by tracing over field states and is given by

$$\rho_{atom}(t)=\sum_{n_1,n_2,\ldots,n_m=0}^{\infty}\left\langle n_1,n_2,\ldots n_m\right|\psi(t)\rangle\langle\psi(t)\left|n_1,n_2,\ldots n_m\right\rangle \tag{11}$$

where, $\left|\psi(t)\right\rangle$ is given by Eq. (7) and, $C$ is called the concurrence defined as

$$C\left(\rho_{atom}\right)=\max\left(0,\sqrt{\lambda_1}-\sqrt{\lambda_2}-\sqrt{\lambda_3}-\sqrt{\lambda_4}\right), \tag{12}$$

where the $\lambda_i$ are the eigenvalues of $\rho_{atom}\left(\sigma_y\otimes\sigma_y\right)\rho^*_{atom}\left(\sigma_y\otimes\sigma_y\right)$ in descending order, and $h$ is the binary entropy function.

## III. Population dynamics and atomic entanglement for a single mode field

Given an initial state, we time evolve the system according to the dynamics governed by Eq. (7) and then trace over the field state.

We use the following derivations

$$\begin{aligned}
&\left(S_+\sum_{k=1}^{m}a_k+S_-\sum_{k=1}^{m}a_k^+\right)^{2p}\\
&=\sum_{i\neq j}\left(|a\rangle\langle b|\right)_i\left(|a\rangle\langle b|\right)_j 2^{p-1}\sum a_k\left(\sum a_k\sum a_k^+ +\sum a_k^+\sum a_k\right)^{p-1}\sum a_k\\
&+\sum_{i\neq j}\left(|b\rangle\langle a|\right)_i\left(|b\rangle\langle a|\right)_j 2^{p-1}\sum a_k^+\left(\sum a_k\sum a_k^+ +\sum a_k^+\sum a_k\right)^{p-1}\sum a_k^+\\
&+\sum_{i\neq j}\left(|a\rangle\langle b|\right)_i\left(|b\rangle\langle a|\right)_j 2^{p-1}\left(\sum a_k\sum a_k^+ +\sum a_k^+\sum a_k\right)^{p}\\
&+\sum_{i\neq j}\left(|a\rangle\langle a|\right)_i\left(|a\rangle\langle a|\right)_j 2^{p-1}\sum a_k\left(\sum a_k\sum a_k^+ +\sum a_k^+\sum a_k\right)^{p-1}\sum a_k^+\\
&+\sum_{i\neq j}\left(|b\rangle\langle b|\right)_i\left(|b\rangle\langle b|\right)_j 2^{p-1}\sum a_k^+\left(\sum a_k\sum a_k^+ +\sum a_k^+\sum a_k\right)^{p-1}\sum a_k\\
&+\sum_{i\neq j}\left(|a\rangle\langle a|\right)_i\left(|b\rangle\langle b|\right)_j 2^{p-1}\left(\sum a_k\sum a_k^+ +\sum a_k^+\sum a_k\right)^{p}
\end{aligned} \tag{13}$$

$$\left(S_+\sum_{k=1}^{m}a_k+S_-\sum_{k=1}^{m}a_k^+\right)^{2p+1}$$

$$=\sum_{i\neq j}\left(|a\rangle\langle a|\right)_i\left(|a\rangle\langle b|\right)_j 2^p\sum a_k\left(\sum a_k\sum a_k^+ +\sum a_k^+\sum a_k\right)^p$$

$$+\sum_{i\neq j}\left(|b\rangle\langle b|\right)_i\left(|a\rangle\langle b|\right)_j 2^p\left(\sum a_k\sum a_k^+ +\sum a_k^+\sum a_k\right)^p\sum a_k$$

$$+\sum_{i\neq j}\left(|a\rangle\langle a|\right)_i\left(|b\rangle\langle a|\right)_j 2^p\left(\sum a_k\sum a_k^+ +\sum a_k^+\sum a_k\right)^p\sum a_k^+$$

$$+\sum_{i\neq j}\left(|b\rangle\langle b|\right)_i\left(|b\rangle\langle a|\right)_j 2^p\sum a_k^+\left(\sum a_k\sum a_k^+ +\sum a_k^+\sum a_k\right)^p \qquad (14)$$

to write the wave function of this atom-field state as

$$|\psi(t)\rangle=|\psi_1(t)\rangle+|\psi_2(t)\rangle+|\psi_3(t)\rangle+|\psi_4(t)\rangle, \qquad (15)$$

where $|\psi_i(t)\rangle$ s' are given as per the situations of the system.

On using the Eq. (6), (7), (8), (13), (14) and (15), we obtain

$$|\psi_1(t)\rangle=\sum_n x_1\ |a_1,a_2,n\rangle \qquad (16)$$

$$|\psi_2(t)\rangle=\sum_n x_2\ |b_1,b_2,n\rangle \qquad (17)$$

$$|\psi_3(t)\rangle=-i\sum_n x_3\ |b_1,a_2,n\rangle \qquad (18)$$

$$|\psi_4(t)\rangle=-i\sum_n x_4\ |a_1,b_2,n\rangle \qquad (19)$$

here, $x_i$ 's are given by

$$x_1=c_{n_k+2}(0)\frac{\sqrt{(n+1)(n+2)}}{2n+3}\left(\cos\left(gt\sqrt{4n+6}\right)-1\right) \qquad (20)$$

$$x_2=c_n(0)\frac{1}{2n-1}\left(n\cos\left(gt\sqrt{4n-2}\right)+n-1\right) \qquad (21)$$

$$x_3=x_4=c_{n+1}(0)\sqrt{\frac{n+1}{4n+2}}\sin\left(gt\sqrt{4n+2}\right) \qquad (22)$$

Hence, the $\rho_{atom}(t)$ of two two-level atoms is expressed as,

$$\rho_{atom}(t)= \sum_{n_1,n_2,\dots,n_m=0}^{\infty} {x_1}^2 |a_1,a_2\rangle\langle a_1,a_2| + \sum_{n_1,n_2,\dots,n_m=0}^{\infty} x_1x_2 |a_1,a_2\rangle\langle b_1,b_2| + i \sum_{n_1,n_2,\dots,n_m=0}^{\infty} x_1x_3 |a_1,a_2\rangle\langle b_1,a_2|$$

$$+ i \sum_{n_1,n_2,\dots,n_m=0}^{\infty} x_1x_4 |a_1,a_2\rangle\langle a_1,b_2| + \sum_{n_1,n_2,\dots,n_m=0}^{\infty} x_2x_1 |b_1,b_2\rangle\langle a_1,a_2| + \sum_{n_1,n_2,\dots,n_m=0}^{\infty} {x_2}^2 |b_1,b_2\rangle\langle b_1,b_2|$$

$$+ i \sum_{n_1,n_2,\dots,n_m=0}^{\infty} x_2x_3 |b_1,b_2\rangle\langle b_1,a_2| + i \sum_{n_1,n_2,\dots,n_m=0}^{\infty} x_2x_4 |b_1,b_2\rangle\langle a_1,b_2| - i \sum_{n_1,n_2,\dots,n_m=0}^{\infty} x_3x_1 |b_1,a_2\rangle\langle a_1,a_2|$$

$$- i \sum_{n_1,n_2,\dots,n_m=0}^{\infty} x_3x_2 |b_1,a_2\rangle\langle b_1,b_2| + \sum_{n_1,n_2,\dots,n_m=0}^{\infty} {x_3}^2 |b_1,a_2\rangle\langle b_1,a_2| + \sum_{n_1,n_2,\dots,n_m=0}^{\infty} x_3x_4 |b_1,a_2\rangle\langle a_1,b_2|$$

$$- i \sum_{n_1,n_2,\dots,n_m=0}^{\infty} x_4x_1 |a_1,b_2\rangle\langle a_1,a_2| - i \sum_{n_1,n_2,\dots,n_m=0}^{\infty} x_4x_2 |a_1,b_2\rangle\langle b_1,b_2| + \sum_{n_1,n_2,\dots,n_m=0}^{\infty} x_4x_3 |a_1,b_2\rangle\langle b_1,a_2|$$

$$+ \sum_{n_1,n_2,\dots,n_m=0}^{\infty} {x_4}^2 |a_1,b_2\rangle\langle a_1,b_2| \quad (23)$$

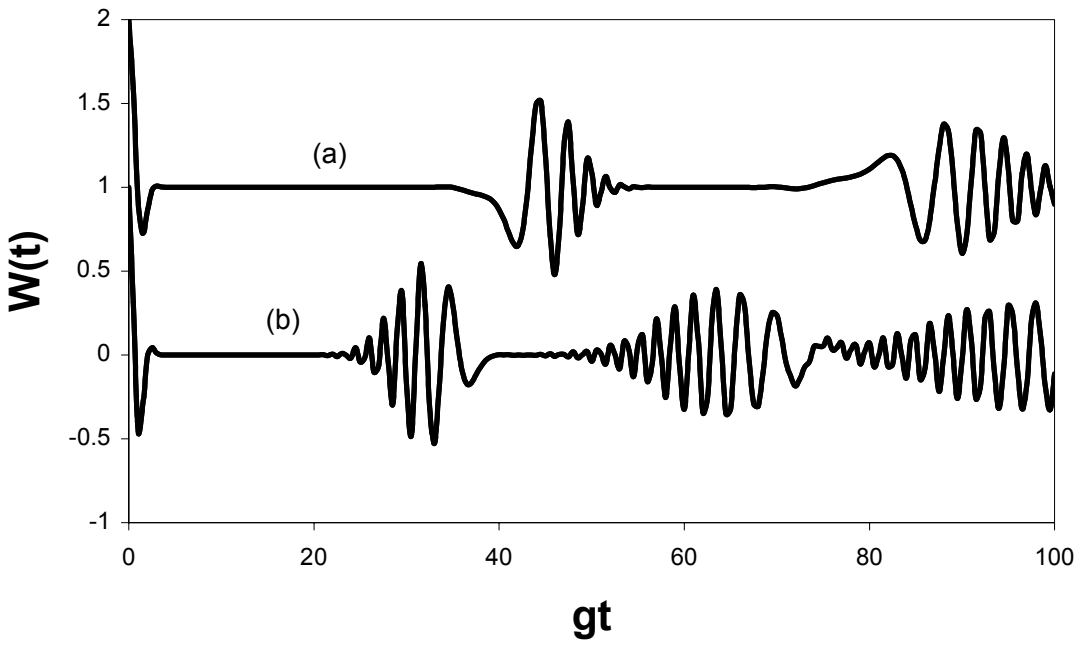

**Figure 1**. Atomic inversion for single atom case (a)<n>=50, (b) <n>=25

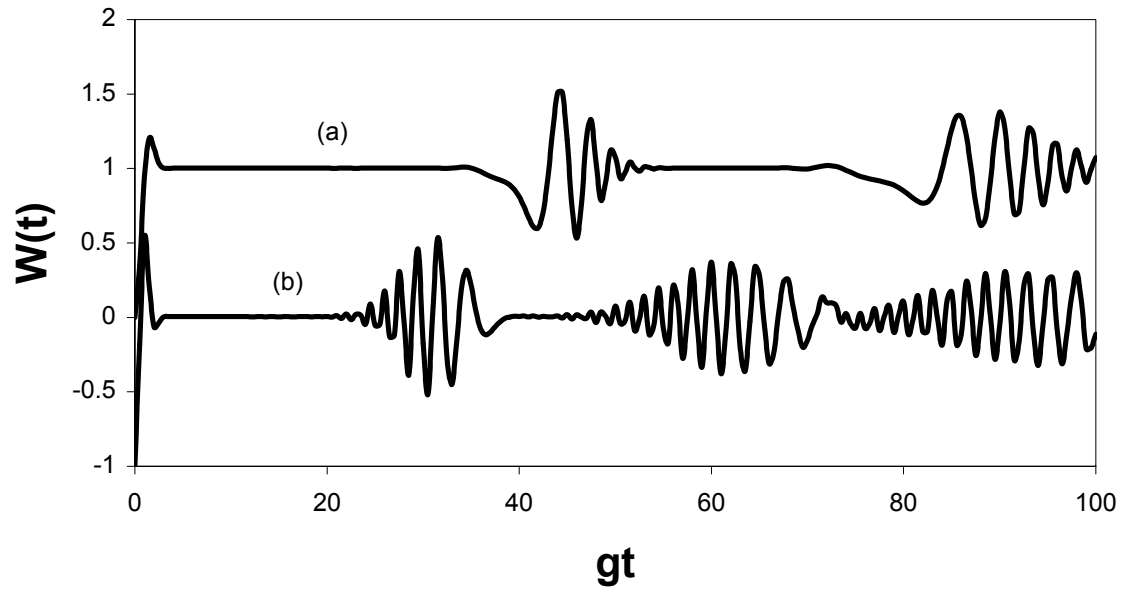

**Figure 2**. Atomic inversion for two atoms (a) <n>=50, (b) <n>=25

In figures (Fig. 1 and Fig. 2) we show the time evolution of the atomic inversion, defined in Eq. (9) as the probability $W$ of finding atom (atoms) in the excited state minus the probability of finding the atom (atoms) in the ground state. We notice in Fig. 1 that the peaks of the revival times agree with the expression $t_j g = 2j\pi\sqrt{\langle n\rangle}$ for the $j^{th}$ revival [20]. We further notice in Fig. 2 that $W(t)$ in the two-atom Tavis-Cummings case follow a similar dynamical pattern. This is due to the initial condition given in Eq.(8) that both atoms in their upper states at $t=0$ [17]. Such a pattern is again reflected in the entanglement properties of the two atoms as described below.

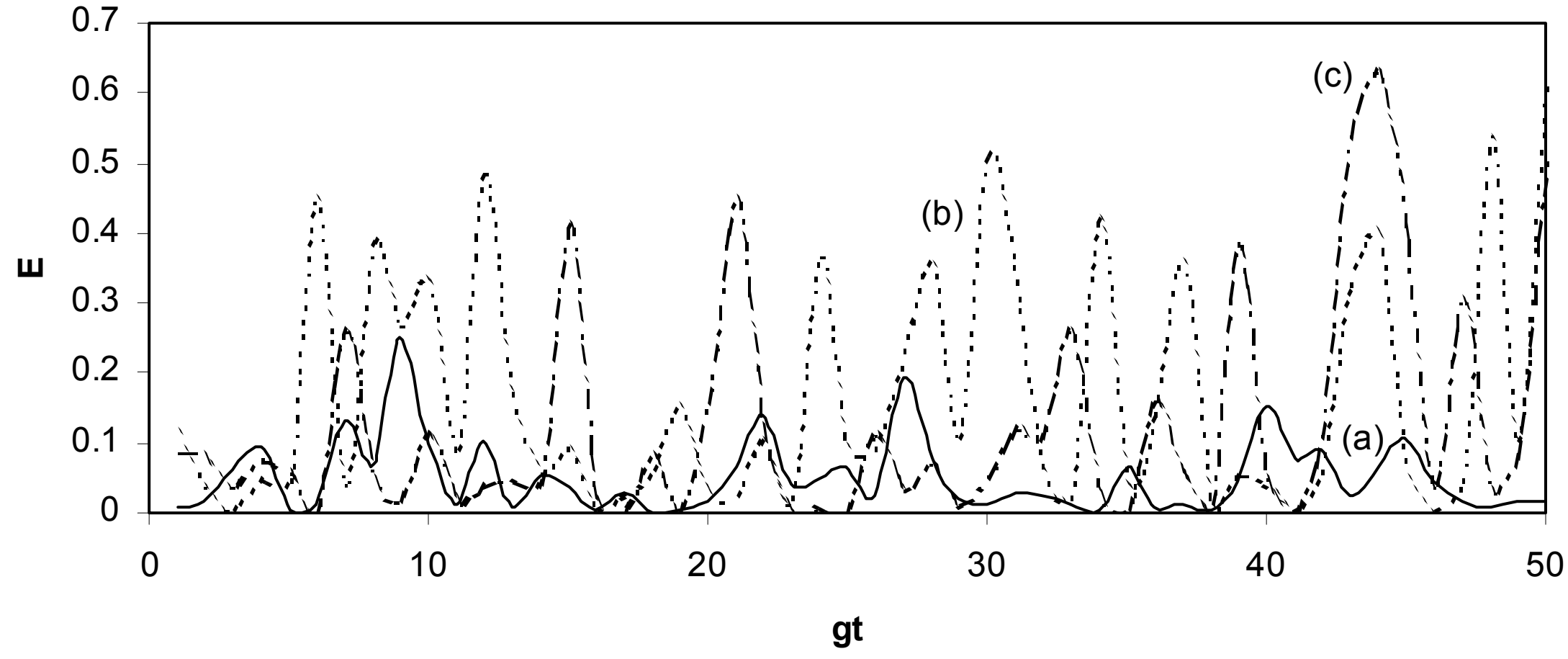

**Figure 3**. Evolution of entanglement for both atoms initially in the excited state and the field (single mode) in an initial coherent state with less photon numbers. (a) solid curve, <n>=0.5; (b) dotted curve, <n>=2.5; and (c) dot-dashed curve, <n>=5.

The entanglement of formation E of the two-atom mixed state $\rho_{atom}(t)$ may then be calculated using Eq.(23) and is plotted in Fig. 3. At certain points of time the entanglement collapses to zero and the two atoms are completely disentangled, while for large time scales revivals of the two atoms entanglement takes place and thus the two atoms behave again as a collective entity, rather than as two individual particles. Gea-Banacloche [21] has shown disentanglement between atom and the field during the course of their interaction in the Jaynes-Cummings model. In the present case, the two atoms interact individually with the common field and in the process the atoms get entangled. The momentarily disentanglement of the two atoms is thus due to the reflection of atom-field disentanglement shown in Ref [21].

## IV. Entanglement of two atoms through multimode ($m > 1$) dynamics

We now consider the general multimode case. Let as make the following assumption in successive calculations,

i.e., $c_{n_k-1}(0) = c_{n_k}(0), c_{n_k+1}(0) = c_{n_k}(0) \forall k, 1 \le k \le m.$ (24)

A coherent state radiation field with high average photon number mostly meets this condition. Our analysis shall center on radiation fields in the above-mentioned conditions.

The wave function for the atom-field system for multimode radiation is given by

$$|\psi_1(t)\rangle = \sum_{n_1, n_2, \ldots, n_m} x_1 \, |a_1, a_2, n_1, n_2, \ldots, n_m\rangle \quad (25)$$

$$\left|\psi_2(t)\right\rangle = \sum_{n_1,n_2,\ldots,n_m} x_2 \left|b_1,b_2,n_1,n_2,\ldots,n_m\right\rangle \tag{26}$$

$$\left|\psi_3(t)\right\rangle = -i \sum_{n_1,n_2,\ldots n_m} x_3 \left|b_1,a_2,n_1,n_2,\ldots,n_m\right\rangle \tag{27}$$

$$\left|\psi_4(t)\right\rangle = -i \sum_{n_1,n_2,\ldots n_m} x_4 \left|a_1,b_2,n_1,n_2,\ldots,n_m\right\rangle \tag{28}$$

where, $x_i$ 's are given by

$$x_1 = \left(\prod_{k=1}^{m} c_{n_k+2}(0)\right) \frac{2\left(\sum_{k=1}^{m}\sqrt{n_k+1}\right)\left(\sum_{k=1}^{m}\sqrt{n_k+2}\right)}{\sum_{\substack{1\le i,j\le m\\ i<j}}\left(\sqrt{n_i+2}+\sqrt{n_j+1}\right)^2+\left(\sqrt{n_i+1}+\sqrt{n_j+2}\right)^2}\times \left(\cos\left(gt\sqrt{\sum_{\substack{1\le i,j\le m\\ i<j}}\left(\sqrt{n_i+2}+\sqrt{n_j+1}\right)^2+\left(\sqrt{n_i+1}+\sqrt{n_j+2}\right)^2}\right)-1\right) \tag{29}$$

$$x_2 = \left(\prod_{k=1}^{m} c_{n_k}(0)\right)\left(\frac{2\left(\sum_{k=1}^{m}\sqrt{n_k}\right)^2}{\sum_{\substack{1\le i,j\le m\\ i<j}}\left(\sqrt{n_i}+\sqrt{n_j-1}\right)^2+\left(\sqrt{n_i-1}+\sqrt{n_j}\right)^2}\times \left(\cos\left(gt\sqrt{\sum_{\substack{1\le i,j\le m\\ i<j}}\left(\sqrt{n_i}+\sqrt{n_j-1}\right)^2+\left(\sqrt{n_i-1}+\sqrt{n_j}\right)^2}\right)-1\right)+1\right) \tag{30}$$

$$x_3 = x_4 = \left(\prod_{k=1}^{m} c_{n_k+1}(0)\right)\frac{\sum_{k=1}^{m}\sqrt{n_k+1}}{\sqrt{\sum_{\substack{1\le i,j\le m\\ i<j}}\left(\sqrt{n_i+1}+\sqrt{n_j}\right)^2+\left(\sqrt{n_i}+\sqrt{n_j+1}\right)^2}}\times \sin\left(gt\sqrt{\sum_{\substack{1\le i,j\le m\\ i<j}}\left(\sqrt{n_i+1}+\sqrt{n_j}\right)^2+\left(\sqrt{n_i}+\sqrt{n_j+1}\right)^2}\right) \tag{31}$$

The $\rho_{atom}(t)$ is defined as per the Eq(23) in the previous section. Unlike Tessier et al. [17], our present investigation can include effect of multimode cavities in atom-atom entanglement. This is important in cases where nearby modes get excited by the interaction of atoms with a particular mode. In fact, confining a cavity in a single mode with high photon number is rather difficult experimentally. We again compute the atomic entanglement by tracing out the field degrees of freedom.

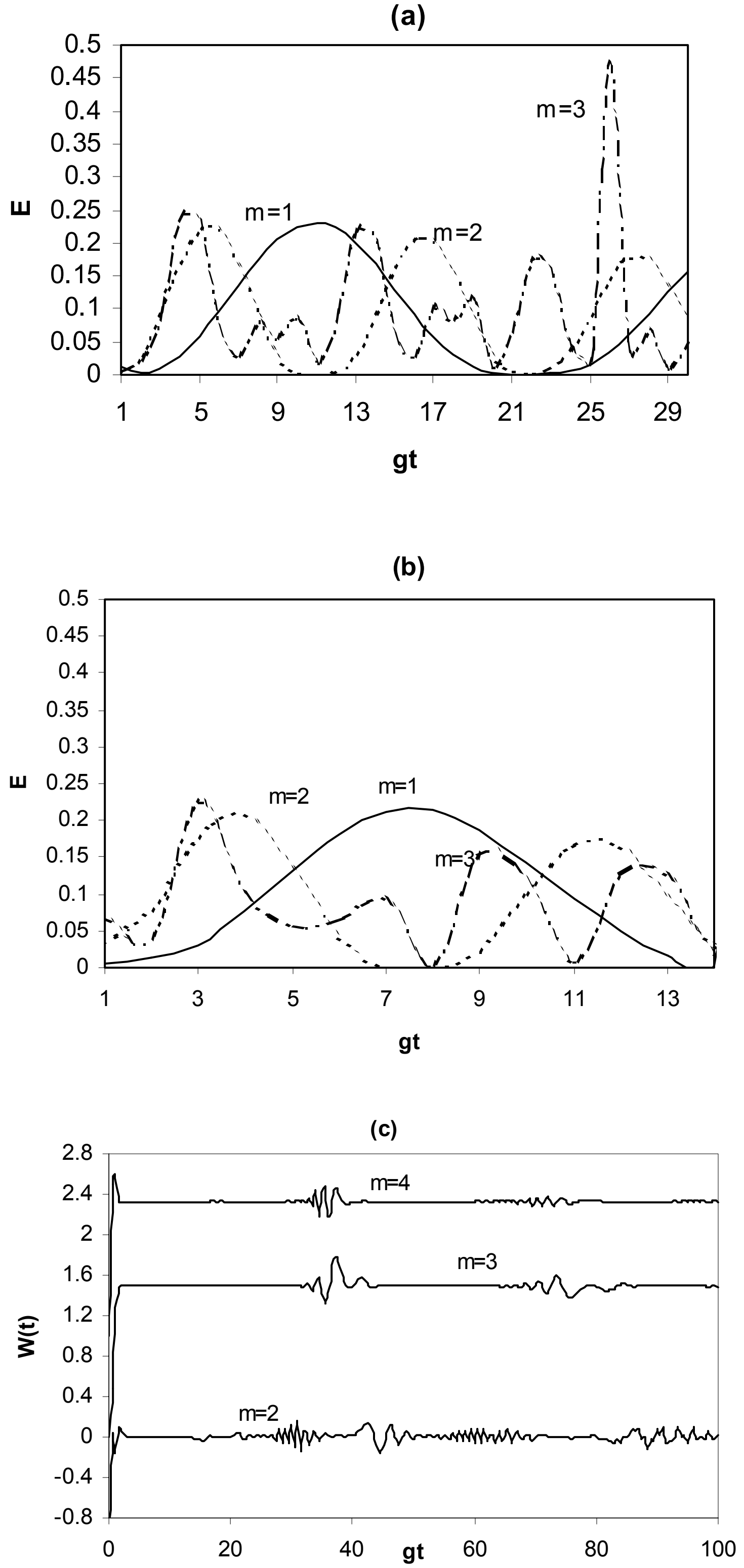

Figure. 4. Time evolution for both atoms initially in the excited states and the field in a coherent state with different values for average photon numbers <n>. For solid curve, m=1; dotted curve, m=2; and dot-dashed curve, m=3. (a) <n>=50. (b) <n>=25. (c) Atomic population dynamics of two-atom m- mode field with <n>=25.

Figure 4 displays the time evolution of the entanglement for two different values of the average photon number. Here, there is no increase in entanglement with increase of field modes. But, it is observed that as the no of modes increases, the frequency of oscillation increases with the modes.

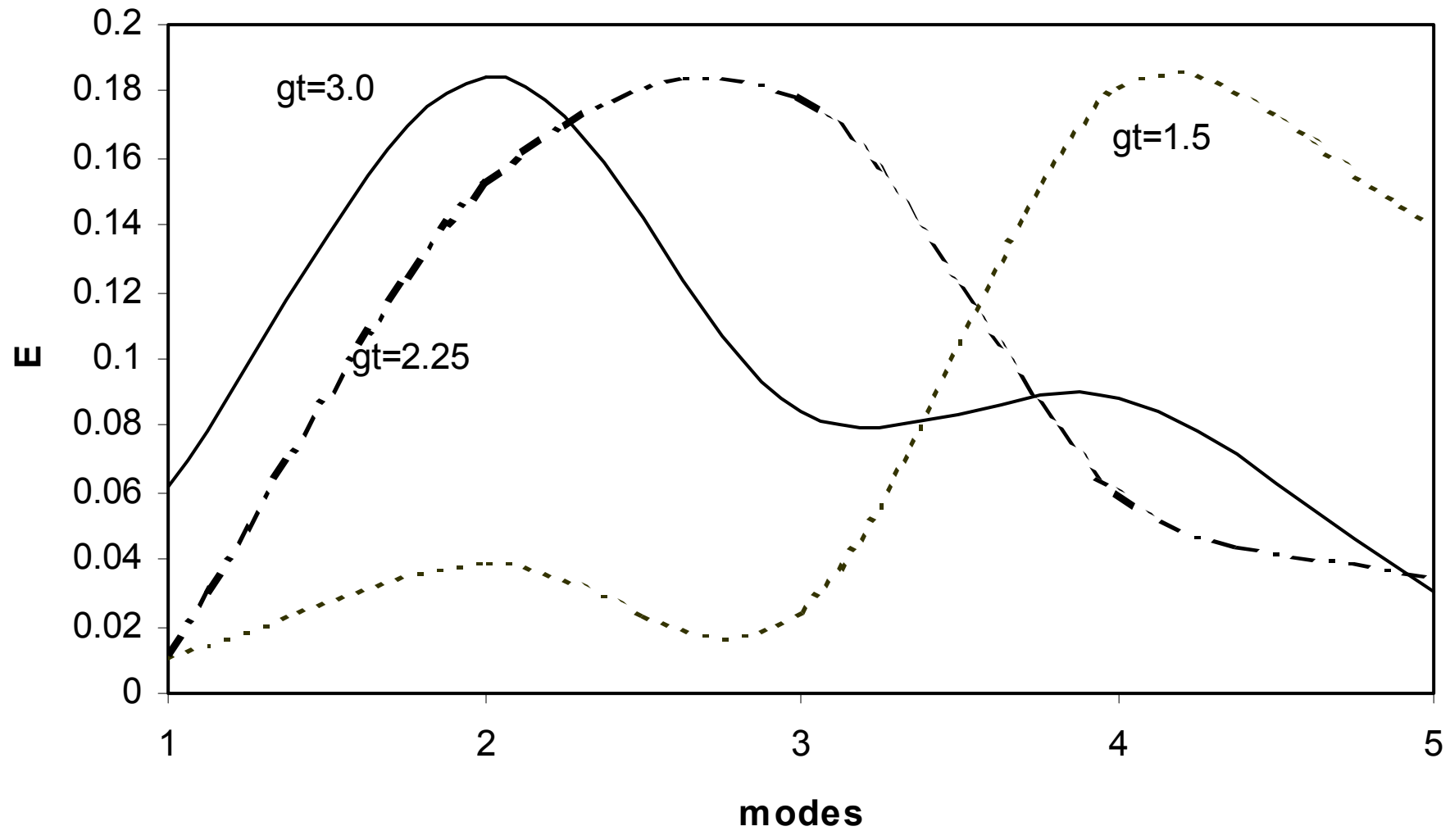

Figure 5. Study of entanglement as a function of no of modes with <n>=15. (a) Solid curve, gt=3.0, (b) dotted curve, gt=1.5, and (c) dot-dashed curve, gt=2.25.

In Fig. 5, we plot the entanglement as the function of modes for different values of gt. This is important from experiments point of view as this figure displays the best entanglement one can get as the number of modes are varied given a particular interaction time. This figure shows that the interaction time *t* is crucial in determining the magnitude of entanglement. Also, the influence of interaction time is different for different modes of the radiation field. This is due to the fact that the oscillations character of the dynamics increases with increase of modes [16]. Here oscillations superimpose in a complicated manner that produce entanglement. Hence, the magnitude of entanglement is bound to vary as the number of modes are increased for a given interaction time. Their effects are important in choosing the parameters in an experimental setup.

## V. Summary

In this paper, we have studied atom-atom (spin-spin) entanglement in a multimode cavity. We consider the two atoms interacting with the modes of the radiation fields. But, we do not consider the direct atom-atom interaction here. Such a system can be achieved by putting the cold atoms in micro cavities. This can also be achieved in a microwave cavity pumped with pairs of Rydberg atoms. The results of our investigations on atom-atom entanglement are presented in the Figs 1-5. We notice in the figures that the atoms get dis-entangled periodically during their time evolution. This can be understood from the fact that, during the *collapse time* in the Jaynes-Cummings model [10], for a single-atom and single-mode interaction, the atom and field get momentarily disentangled. In the present case both the atoms know *each other* only through the field. When the two atoms get disentangled from the fields with which they are interacting, it makes one atom forget the other. Comparisons with population dynamics shows that the disentanglement takes place in the middle of collapse time.

We have considered an ideal cavity, that is, the cavity losses are assumed to be zero. But, we have shown in an earlier work [22] that the reservoir, representing the loss of cavity photons, could play a major role in atom-atom entanglement. The reservoir can destroy atom-atom entanglement and, also, at times can help in building up atom-atom entanglement. Thus a complete picture of the two atoms interacting with a multimode cavity and its reservoir can show that the entanglement is lost to the reservoir only to be regained at the later time. The collapse and revival of entanglement of two qubits in the presence of a reservoir has been recently studied [23]. We plan to further investigate this issue within the context of the Tavis Cummings model in a subsequent work.

**References**

[1] Plenio M B, Huelga S F, Beige A and Knight P L 1999 Phys. Rev. A **59** 2468
[2] Ficek Z and Tanas R 2002 Phys. Rep. **372** 369
[3] Malinovsky V S and Sola I R 2004 Phys. Rev. Lett. **93** 190502
[4] Serafini A, Mancini S, and Bose S 2006 Phys. Rev. Lett. **96** 010503
[5] Polzik E S 1999 Phys. Rev A **59** 4202
[6] Zyczkowski K, Horedecki P, Horodecki M and Horodecki R 2001 Phys. Rev. A **65** 012101
[7] Slichter C P 1978 in *Principles of Magnetic Resonance*, Springer-Verlag, Berlin
[8] Berman P R 1994 in *Cavity Quantum Electrodynamics*, Academic, New York
[9] Nielson M A and Chuang I L 2001 *in Quantum Computation and Quantum Information*, Cambridge University Press, Cambridge
[10] Jaynes E T and Cummings F W 1963 Proc IEEE **51** 89
[11] Tavis M and Cummings F W 1968 Phys. Rev. **170** 379
[12] Genes C, Berman P R and Rojo A G 2003 Phys. Rev. A **68** 043809
[13] Haroche S and Raimand J M 2001 Rev. Mod. Phys **73** 565
[14] Kim M S, Lee J -H, Ahn D and Knight P L 2002 Phys Rev. A **65** 040101
[15] Ghosh B, Majumdar A S and Nayak N 2006 Int. J. Quant. Inf. **4** 665
[16] Saha P 2007 Phys. Rev. A **76** 023804
[17]Tessier T E, Deutsch I H and Delgado A 2003 Phys Rev A **68** 062316
[18] Hill S and Wootters W K 1997 Phys. Rev. Lett. **78** 5022
[19] Wootters W K 1998 Phys. Rev. Lett. **80** 2245

[20] Eberly J H, Narozhny N B and Sanchez-Mondragan J J 1980 Phys. Rev. Lett. **44** 1323; Nayak N, Bullagh R K, Thompson B V and Agarwal G S 1988 IEEE J. Quant. Electron. **QE-24** 1331
[21] Gea-Banacloche J 1990 Phys. Rev. Lett. **65** 3385
[22] Ghosh B, Majumdar A S and Nayak N 2006 Phys. Rev. A **74** 052315
[23] Das S and Agarwal G S 2009 arXiv:**0901.2114**